\documentclass[times, 10pt,twocolumn]{article}
\usepackage{graphicx}
\usepackage{latex8}
\usepackage{times}

\begin{document}
\title{Applications of Cluster Perturbation Theory Using Quantum Monte Carlo Data}
\author{Fei Lin, Erik S. S$\o$rensen, Catherine Kallin and A. John Berlinsky}
 \affiliation{Department of Physics and Astronomy, McMaster University, Hamilton, Ontario,
 Canada L8S 4M1}

\maketitle
\thispagestyle{empty}

\begin{abstract}
We study cluster perturbation theory [Phys. Rev. Lett.
\textbf{84}, 522 (2000)] when auxiliary field quantum Monte Carlo
method is used for solving the cluster hamiltonian. As a case
study, we calculate the spectral functions of the Hubbard model in
one and two dimensions and compare our results for the spectral
functions to results obtained using exact diagonalization to solve
the cluster hamiltonian. The main advantage of using quantum Monte
Carlo results as a starting point is that the initial cluster size
can be taken to be considerably larger and hence potentially
capture more of the relevant physics. The drawback is that quantum
Monte Carlo methods yield results at {\it imaginary} times with
stochastic errors.
\end{abstract}

\section{Introduction}
In the field of strongly-correlated systems (such as the
high-temperature superconductors), it is usually useful to study
the single-particle excitation spectrum. Experimentally, it can be
obtained through the ARPES technique for some systems
\cite{shen03}; while theoretically, it can be calculated by exact
diagonalization (ED) or quantum Monte Carlo (QMC) method for some
finite lattice models, such as the Hubbard model, which describes
a lattice system of strongly-interacting electrons and which is
believed by some Physicists to capture the Physics of the
high-temperature superconductors.

Since these calculations are for a finite system, it is desirable
to extend the calculations to the infinite lattice systems for a
better comparison between the experiments and the theoretical
calculations. Cluster perturbation theory (CPT) as proposed by
S$\acute{e}$n$\acute{e}$chal \textit{et al.} \cite{senechal00,
senechal02} is one of these extensions, and has turned out to be
an accurate and economical technique for calculating the spectral
functions of the Hubbard model in one- and two-dimensional (1D and
2D) lattice systems. The method starts by dividing the infinite
lattice system into a periodic array of clusters. The Hubbard
model inside a cluster is then solved by ED, and the hopping
matrix between two neighboring clusters is treated as a
perturbation. The cluster Green's functions from ED are then used
to perturbatively construct the Green's functions of the infinite
lattice. The method is economical in that one usually only
diagonalizes a small cluster (typically less than 14 sites for the
Hubbard model), and the infinite lattice Green's function with an
arbitrary momentum $\textbf{k}$ can then be constructed with a
very small computational overhead. The computationally limiting
step is therefore the exact diagonalization of the cluster which
effectively limits its size to less than 20 sites or so for the
Hubbard model. (The computer memory requirement grows
exponentially with increasing cluster sizes.) On the other hand
many materials, such as molecular solids, display a ``natural"
cluster size (the molecule) which may be significantly beyond what
can be treated with exact diagonalization techniques and it
therefore becomes of great interest to study the accuracy of
cluster perturbation theory when the cluster itself is treated
with approximate techniques.

In this paper we study the accuracy of the CPT method
\cite{senechal00, senechal02} when auxiliary field quantum Monte
Carlo (AFQMC) methods \cite{hirsch85, hirsch88, white89} are used
to solve the cluster Hamiltonian. To distinguish between these two
methods, we call the usual CPT method using exact diagonalization
methods to solve the cluster hamiltonian EDCPT and analogously the
present method QMCPT. The advantage of QMCPT is that we can deal
with clusters as large as 60 sites (results will be presented
elsewhere) as long as the sign problem \cite{hirsch85}, which is
the appearance of negative probabilities in the QMC simulations,
is not too severe. However, AFQMC yields {\it imaginary} time
Green's functions with stochastic errors. In order to use such
Green's functions as input to the zero temperature cluster
perturbation theory formalism an analytic continuation to real
times (frequency) has to be performed, a notoriously difficult
step. Hence, it is not obvious that this approach will yield
reliable results. Fortunately, as we show in the following, if
high precision numerical data are available for the imaginary time
Green's functions an analytical continuation using maximum entropy
methods of the AFQMC data yield results that are in quite good
agreement with EDCPT for the same cluster size.

In the next section we briefly introduce the EDCPT method. This is
followed by the introduction of AFQMC and QMCPT, describing
details in the calculation. Spectral functions of the 1D and 2D
Hubbard Hamiltonians are then presented and compared to those
obtained using EDCPT.

\section{Model}
The model Hamiltonian we consider is given by
\begin{eqnarray}
H&=&H_0+V,\\
H_0&=&\sum_I H_0^I,
\end{eqnarray}
where
\begin{equation}
H_0^I=-t\sum_{\langle ij\rangle
\sigma}^I(c^{\dagger}_{i\sigma}c_{j\sigma}+h.c.)+U\sum_i^I
n_{i\uparrow}n_{i\downarrow} \label{cterm}
\end{equation}
is the Hubbard model inside the $I$'th cluster, and
\begin{equation}
V=-t^{'}\sum_{\langle Ii,Jj \rangle
\sigma}(c^{\dagger}_{Ii\sigma}c_{Jj\sigma}+h.c.) \label{Vterm}
\end{equation}
is the hopping terms between two nearest neighbor (NN) sites $i$
and $j$ in two NN clusters $I$ and $J$, respectively. Here the
operator notations are standard for the usual Hubbard model, i.e.,
$c^{\dagger}_{i\sigma} (c_{i\sigma})$ is the electron creation
(annihilation) operator for spin $\sigma$, and $n_{i\sigma}$ is
the electron number operator for site $i$ with spin $\sigma$.
Inside the cluster $U$ is the on-site Coulomb interaction energy,
and $t$ is the hopping integral between two NN sites. When
$t^{'}=t$, we recover the standard Hubbard model. In the following
calculations $t$ is used as the energy unit.

\begin{figure}
  \begin{tabular}{c}
  \resizebox{80mm}{!}{\includegraphics{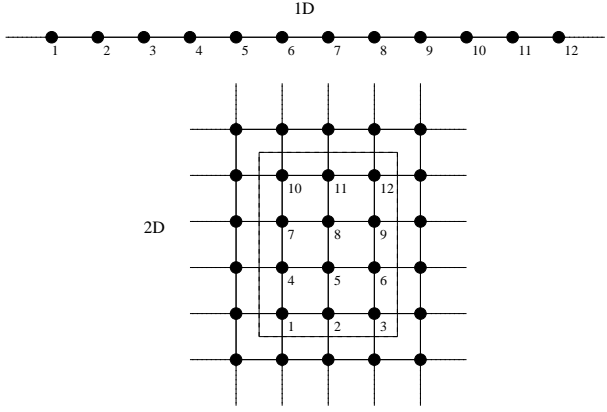}} \\
  \end{tabular}
  \caption{1D and 2D clusters for the CPT calculations. In 1D a
  cluster of 12 sites is used in the paper. A $3\times 4$ cluster
  is drawn for the 2D square lattice.}
  \label{1d2d}
\end{figure}
\begin{figure}
  \begin{tabular}{c}
  \resizebox{80mm}{!}{\includegraphics{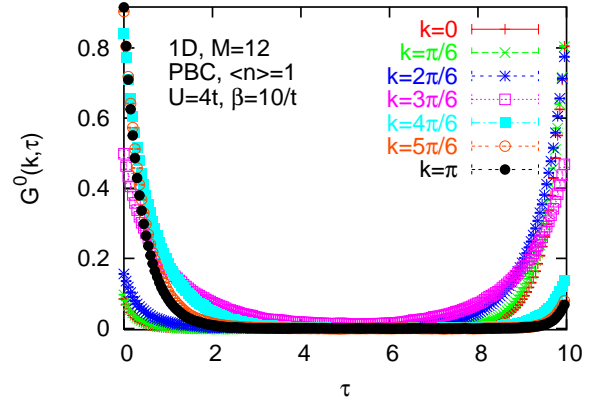}} \\
  \end{tabular}
  \caption{Evolution of cluster Green's function with imaginary
  time $\tau$ in QMC simulation. We see that enough amount of data has made
  the error bars very small, which is essential in MEM \cite{jarrell96}.}
  \label{gtau}
\end{figure}
\section{EDCPT Formalism}
To compare EDCPT with QMCPT, we will briefly describe the
calculational steps of EDCPT for the Hubbard model. EDCPT first
uses the Lanczos algorithm \cite{golub89}, which is an efficient
method for obtaining several lowest eigen values of sparse
matrices, to diagonalize the Hubbard Hamiltonian $H_0^I$ of a
cluster for its ground state energy $E_0$ and the corresponding
eigen state $|\Psi\rangle$. The single-particle Green's function
is then given by
\begin{eqnarray}
G_{ij}^0(\omega)&=&G_{ij}^e(\omega)+G_{ij}^h(\omega),\label{g0define}\\
G_{ij}^e(\omega)&=&\langle\Psi|c_i\frac{1}{\omega+i\eta+E_0-H_0^I}
c_j^{\dagger}|\Psi\rangle,\\
G_{ij}^h(\omega)&=&\langle\Psi|c_i^{\dagger}\frac{1}{\omega+i\eta-E_0+H_0^I}
c_j|\Psi\rangle,
\end{eqnarray}
where $e$ and $h$ represent respectively the electron and hole
part of the green's function, and $\eta$ is a small positive
parameter to give the Lorenzian broadening of the delta functions.
The calculation of these cluster Green's functions is standard,
and can be found in the literature (e.g., Ref.\ \cite{lin93}).

With these Green's functions $G^0_{ij}(\omega)$,
the Green's functions for the superlattice $G(\textbf{K},\omega)$
is constructed through the strong coupling perturbation \cite{senechal00, senechal02}:
\begin{equation}
G_{ij}(\textbf{K},\omega)=\left (\frac{G^0(\omega)}
{1-V(\textbf{K})G^0(\omega)}\right )_{ij}, \label{gk1}
\end{equation}
where $\textbf{K}$ is superlattice's momentum vector, and
$V(\textbf{K})$ is the Fourier transform of Eq.\ (\ref{Vterm}).
$V(\textbf{K})$ is given by
\begin{equation}
V_{ij}(\textbf{K})=\sum_{\textbf{R}}V_{ij}^{0,\textbf{R}}e^{i\textbf{K}\cdot\textbf{R}},
\label{vk}
\end{equation}
where $\textbf{R}$ represents position of the cluster in the
superlattice. A residual Fourier transform is performed to
construct the lattice Green's functions in terms of the
$\textbf{k}$ vectors of the original lattice
\begin{equation}
G_{\textrm{CPT}}(\textbf{k},\omega)=\frac{1}{M}\sum_{i,j=1}^M
G_{ij}(\textbf{k},\omega)e^{-i\textbf{k}\cdot(\textbf{r}_i-\textbf{r}_j)},
\label{gk2}
\end{equation}
where $M$ is the number of lattice sites in one cluster. The
spectral function of the infinite lattice is then given by
\begin{equation}
A(\textbf{k},\omega)=-\frac{1}{\pi}\lim_{\eta\rightarrow 0^{+}}
\textrm{Im}G_{\textrm{CPT}}(\textbf{k},\omega). \label{akw}
\end{equation}

\begin{figure}
  \begin{tabular}{c}
  \resizebox{70mm}{!}{\includegraphics{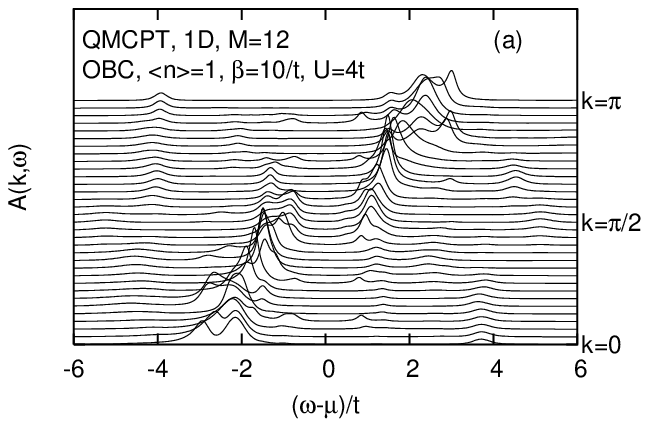}} \\
  \resizebox{70mm}{!}{\includegraphics{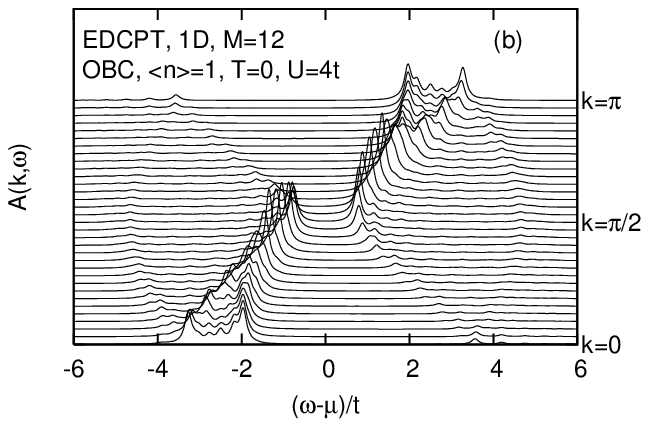}} \\
  \end{tabular}
  \caption{Single particle spectral functions of the 1D Hubbard
  model from (a) QMCPT and (b) EDCPT with open boundary conditions.
  Both methods produce roughly the same single-particle excitation
  energies $(\omega-\mu)/t$ and energy gap value, although the spectral
  height is different at some energy values. See text for the reason.}
  \label{1du4obc}
\end{figure}
\begin{figure}
  \begin{tabular}{c}
  \resizebox{70mm}{!}{\includegraphics{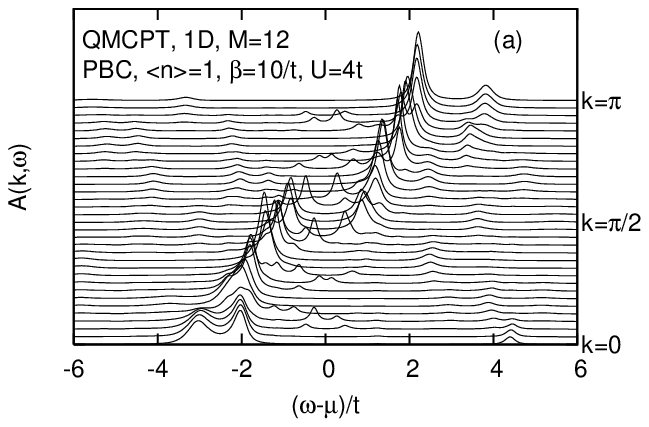}} \\
  \resizebox{70mm}{!}{\includegraphics{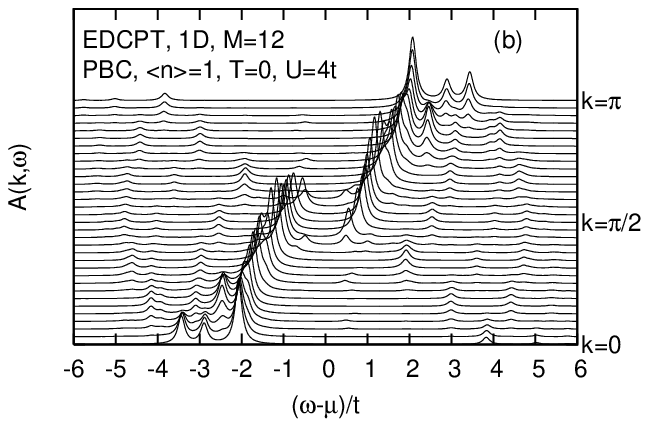}} \\
  \end{tabular}
  \caption{Comparison of single particle spectral functions of a 1D Hubbard
  model from (a) QMCPT and (b) EDCPT with periodic boundary conditions.
  Both methods give roughly the same single-particle excitation energies
  at most energy values, but the spurious excitations inside the energy
  gap shows that QMCPT with periodic boundary conditions is inaccurate.}
  \label{1du4pbc}
\end{figure}
\begin{figure}
  \begin{tabular}{c}
  \resizebox{70mm}{!}{\includegraphics{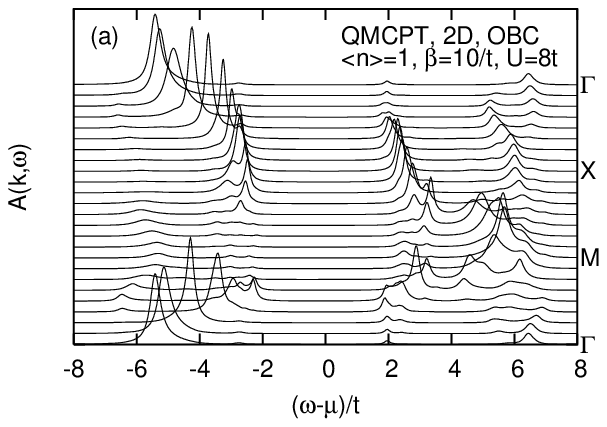}} \\
  \resizebox{70mm}{!}{\includegraphics{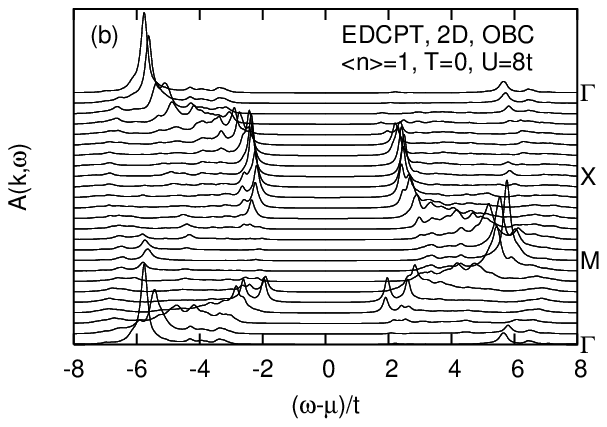}}\\
  \end{tabular}
  \caption{Single particle spectral functions of a 2D Hubbard model
  from (a) QMCPT and (b) EDCPT with open boundary conditions.
  The cluster is of dimension $3\times 4$. Both methods predict the same
  single-particle excitation energies and energy gap, though at some energy
  values the spectral height is different. See text for the reason.}
  \label{2du8}
\end{figure}
\section{QMCPT Formalism}
In QMCPT the imaginary-time cluster Green's functions
$G^0(\textbf{k},\tau)$ are calculated using the auxiliary field
QMC technique \cite{hirsch85, hirsch88, white89}. They are then
analytically continued to extract the real frequency spectral
functions $A^0(\textbf{k},\omega)$ through the Maximum Entropy
method (MEM) \cite{jarrell96}:
\begin{equation}
G^0(\textbf{k},\tau)=\int
d\omega\frac{e^{-\tau\omega}A^0(\textbf{k},\omega)}{1+e^{-\beta\omega}},
\label{memg}
\end{equation}
where $\beta=1/k_{\textrm{B}}T$ is the inverse temperature in the
QMC simulation, and $\tau$ is the imaginary time. The above
integral in Eq.\ (\ref{memg}) is discretized with a small
$\Delta\omega$ and summed over a set of $A^0(\textbf{k},\omega)$
values. A smaller $\Delta\omega$ produces a smoother image for
$A^0(\textbf{k},\omega)$, but different $\Delta\omega$ usually
gives results in agreement with each other. We refer the
interested readers to Ref.\ \cite{jarrell96} for the details of
MEM. The resulting $A^0(\textbf{k},\omega)$ can be used to
construct the real frequency cluster Green's functions via
\cite{fetter03}
\begin{equation}
G^0(\textbf{k},\omega)=\int
d\omega^{'}\frac{A^0(\textbf{k},\omega^{'})}{\omega+i\eta-\omega^{'}}.
\end{equation}
After a cluster Fourier transform from the $\textbf{k}$ vectors to
cluster $i,j$ indices, and again following Eqs.\ (\ref{gk1}),
(\ref{vk}), (\ref{gk2}) and (\ref{akw}), we can evaluate the
spectral functions $A(\textbf{k},\omega)$ for the infinite
lattice.

Next we discuss in detail the calculation of space-time Green's
functions with QMC simulations. Since we are considering only the
cluster Hamiltonian, we can neglect Eq.\ (\ref{Vterm}) and write
Eq.\ (\ref{cterm}) as
\begin{equation}
H_0^{I}=T+W,
\end{equation}
where
\begin{equation}
T=-t\sum_{\langle ij\rangle
\sigma}(c^{\dagger}_{i\sigma}c_{j\sigma}+h.c.)
\end{equation}
and
\begin{equation}
W=U\sum_i n_{i\uparrow}n_{i\downarrow}
\end{equation}
representing, respectively, the kinetic and potential energies of
the cluster.

The partition function is then given by
\begin{eqnarray}
Z&=&\textrm{Tr}\,e^{-\beta(T+W)}\nonumber\\
&=&\textrm{Tr}\,\prod_{i=1}^{L}e^{-\Delta\tau(T+W)},
\end{eqnarray}
where $\beta=\Delta\tau L$, and $L$ is the number of time slices
in the imaginary time direction. After tracing out the fermionic
operators, we get
\begin{eqnarray}
Z&=&\sum_{\{\sigma\}}\prod_{\alpha}\det
[1+B_{L}(\alpha)B_{L-1}(\alpha)\cdots
B_{1}(\alpha)]\nonumber\\
&=&\sum_{\{\sigma\}}\det O(\{\sigma\},\mu)_{\uparrow}\det
O(\{\sigma\},\mu)_{\downarrow}. \label{zfinal}
\end{eqnarray}

The $B_l$ matrices are defined as
\begin{eqnarray}
B_{l}(\alpha)&=&e^{-\Delta\tau T}e^{W^{\alpha}(l)},\label{bdefine}\\
(T)_{ij}&=&\left\{\begin{array}{cc}
-t_{ij} & \mbox{for $i$,$j$ NN},\\
0 & \mbox{otherwise},
\end{array}\right.\\
W_{ij}^{\alpha}(l)&=&\delta_{ij}[\lambda\alpha\sigma_{i}(l)+\mu\Delta\tau],
\label{vls}
\end{eqnarray}
where $\mu$ is chemical potential, $\tanh^2(\lambda/2)=\tanh(\Delta\tau U/4)$,
$\sigma_i(l)=\pm 1$ is the auxiliary Ising spin coupled with the electrons at lattice site
$i$ and time $\tau=(l-1)\Delta\tau$, and $\alpha=\pm 1$ corresponds to
$\uparrow$ or $\downarrow$ in Eq.\ (\ref{zfinal}).

In calculating the space-time Green's functions, we form a matrix
of dimension $ML\times ML$ after a complete sweep over the $M$
cluster sites and $L$ time slices, the inverse of which gives the
desired cluster Green's functions \cite{assaad02, hirsch88}
\begin{eqnarray}
& &\left(\begin{array}{ccccc}
1 & 0 & \cdot & 0 & B_1 \nonumber \\
-B_2 & 1 & 0 & \cdot & 0 \nonumber \\
0 & -B_3 & 1 & \cdot & 0 \nonumber \\
\cdot & 0 & -B_4 & \cdot & \cdot \nonumber \\
\cdot & \cdot & 0 & \cdot & \cdot \nonumber \\
\cdot & \cdot & \cdot & \cdot & 0 \nonumber \\
0 & \cdot & 0 & -B_L & 1 \nonumber
\end{array}\right)^{-1}=\nonumber \\
& &\left(\begin{array}{ccccc}
G^0(1,1) & G^0(1,2) & \cdot & \cdot & G^0(1,L) \nonumber\\
G^0(2,1) & G^0(2,2) & \cdot & \cdot & G^0(2,L) \nonumber\\
\cdot & \cdot & \cdot & \cdot & \cdot \nonumber \\
G^0(L,1) & G^0(L,2) & \cdot & \cdot & G^0(L,L) \nonumber\\
\end{array}\right),
\end{eqnarray}
where, for $l_1>l_2$,
\begin{eqnarray}
G_{ij}^0(l_1,l_2)&=&\langle
c_i(l_1)c_j^{\dagger}(l_2)\rangle\nonumber\\
&=&[B_{l_1}B_{l_1-1}\cdots B_{l_2+1}(1+\nonumber\\
&&B_{l_2}\cdots B_1B_L\cdots B_{l_1+1})^{-1}]_{ij}.\label{gdefine}
\end{eqnarray}
After a Fourier transform of the above Green's functions, we have
$G_{\textbf{k}}^0(l,1)=G^0(\textbf{k},\tau)$, where
$\tau=(l-1)\Delta\tau$ with $(1\leq l\leq L)$. They are the input
for the MEM algorithm in Eq.\ (\ref{memg}). Note that in the above
definition of finite temperature Green's functions in Eq.\
(\ref{gdefine}) we used the same symbol as that for zero
temperature in Eq.\ (\ref{g0define}), since there is no difference
in the application of the CPT method in these two cases.

\section{Results}
\subsection{1D Case}
In this section we will present QMCPT results on the 1D and 2D
Hubbard models. See Fig.\ \ref{1d2d} for the division of the 1D
and 2D lattices into clusters and the geometry of the respective
clusters. We choose 12-site clusters in both 1D and 2D cases, so
that we can compare the results with those available in the
literature. The intra-cluster hopping integral $t$ is used as the
energy unit, and the on-site Coulomb interaction $U=4t$ or $U=8t$.
The inter-cluster hopping integral $t^{'}$ will be set to $t$,
too. In QMC the inverse temperature is $\beta=10/t$, which will
produce results close to the ground state. The Lorenzian
broadening parameter is $\eta=0.1t$, which is of the same order of
the QMC temperature.

In QMC simulations we discretize the imaginary time $\beta=10/t$
into $200$ slices, i.e., $\Delta\tau=0.05/t$ to make the Trotter
error smaller than the statistical ones. The 12-site Hubbard ring
assumes either open boundary conditions (OBC) or periodic boundary
conditions (PBC). One thousand complete sweeps over the space-time
lattice are performed to warm up the system. For each boundary
case, we collect $51,000$ sets of space-time Green's functions,
each $100$ sets of which are averaged as 1 bin (total $510$ bins)
for the subsequent MEM analysis. A typical $G^0(\textbf{k},\tau)$
versus $\tau$ is shown in Fig.\ \ref{gtau}. For generating a
smooth figure for $A^0(\textbf{k},\omega)$, the discretization of
the real frequency is set at $\Delta\omega=0.025t$ in MEM. Note
that a smaller $\Delta\omega$ does not change the resultant figure
except for making the figure smoother with longer computation
times. Fig.\ \ref{1du4obc} compares the QMCPT result with the
EDCPT result for a 1D Hubbard chain constructed from the 12-site
cluster with open boundary conditions. We find that QMCPT result
agree very well with the EDCPT one. They both have a Hubbard gap
opening at $\omega-\mu=0$ and the same positions of spectral
peaks, although the peak heights are different in some positions.
Hence, a precise determination of the quasi-particle weights is
probably not feasible with QMCPT. We expect this to happen,
because MEM always broadens the sharp peaks or blurs the fine
details of the exact results. In spite of this problem, QMCPT
still produces spectral functions that satisfy the usual sum rules
\cite{fetter03}.

There has been a suggestion of using periodic boundary condition (PBC)
in the clusters \cite{hanke02}, which, in ED, can greatly
reduce the dimension of the Hilbert space through the
translational symmetry. These additional hopping terms are then
subtracted from the perturbation $V$ during the construction of
the superlattice Green's functions. Comparison of the EDCPT
results with open and periodic boundary conditions are made in Ref.\ \cite{senechal02},
where both results agree in the sense that a Hubbard gap opens at
half filling for both cases. Here we repeat the same calculations
in the same systems with both QMCPT and EDCPT. The results are shown in
both Fig.\ \ref{1du4obc} and Fig.\ \ref{1du4pbc}. We find that QMCPT
calculations with periodic boundary conditions produce spurious
single particle peaks around $\omega-\mu=0$. Therefore, in what
follows we use open boundary conditions in the QMCPT calculations.

\subsection{2D Case}
In the 2D Hubbard system, we choose a cluster of $M_xM_y=3\times
4$ (see Fig.\ \ref{1d2d}) with open boundary conditions for QMC
simulation. We have collected $71,000$ sets of space-time Green's
functions, which are averaged to yield $710$ bins for MEM
analysis. We find that in this $U=8t$ case less sets of data makes
the MEM analysis difficult \cite{jarrell96}. The QMCPT result
together with the EDCPT result are shown in Fig.\ \ref{2du8},
where the $\textbf{k}$ vector scan is along $\Gamma-M-X-\Gamma$
(Please see the figure illustration of this momentum space scan in
Ref.\ \cite{senechal00}). We see that the QMCPT produces results
close to those of EDCPT. (See also Ref.\ \cite{senechal00}.) The
main peaks found in EDCPT are all preserved in the QMCPT
calculation, which again shows that we can safely use QMCPT for
the lattice spectral function calculations (in that QMCPT predicts
the correct single-particle excitation energies and energy gaps
though with different spectral heights from EDCPT at some energy
values). This is again because of the broadening of sharp peaks
and the blurring of fine details in the MEM analysis.

\section{Conclusions}
In summary, we have presented our first effort in applying CPT
method using AFQMC to solve the cluster hamiltonian. At low temperatures our QMC and QMCPT
results agree very well with those from EDCPT
\cite{senechal00,senechal02}. Compared with EDCPT, QMCPT works at
finite and very low temperatures, and the cluster sizes can be much larger. We
expect QMCPT to be a useful tool in calculating the spectral
functions in not only the 1D and 2D lattices but also some
molecular solids, e.g., fullerene materials, where each molecule
naturally defines a cluster, and ED of Hubbard model is not
possible for molecules with more than 20 sites.

\subsection*{Acknowledgements}

This project was supported by the Natural Sciences and Engineering
Research Council (NSERC) of Canada, the Canadian Institute for
Advanced Research (CIAR), and the Canadian Foundation for
Innovation (CFI). All the calculations were carried out at
SHARCNET supercomputing facilities at McMaster University.

\bibliographystyle{latex8}
\bibliography{qmcpt_hpc06}

\end{document}